# Assembling of three-dimensional crystals by large nonequilibrium depletion force

Hai-Dong Deng,[1] Ting Sun,[1] Zhi-Cheng Fu,[1] Hai-Ying Liu,[1] Qiao-Feng Dai,[1] Li-Jun Wu,[1] Sheng Lan,[1,a)] and Achanta Venu Gopal[2]

[1]*Laboratory of Photonic Information Technology, School for Information and Optoelectronic Science and Engineering, South China Normal University, Guangzhou 510006, P. R. China*

[2]*Department of Condensed Matter Physics and Material Science, Tata Institute of Fundamental Research, Homi Bhabha Road, Mumbai 400005, India*

**Abstract**

We propose and demonstrate a method to achieve large effective Soret coefficient in colloids by suitably mixing two different particles, e.g., silica beads and $Fe_3O_4$ nanoparticles. It is shown that the thermophoretic motion of $Fe_3O_4$ nanoparticles out of the heating region results in a large nonequlibrium depletion force for silica beads. Consequently, silica beads are driven quickly to the heating region, forming a three-dimensional crystal with few defects and dislocations. The binding of silica beads is so tight that a colloidal photonic crystal can be achieved after the complete evaporation of solvent, water. Thus, for fabrication of defect free colloidal PCs, periodic structures for molecular sieves, among others, the proposed technique could be a low cost alternative. In addition as we use biocompatible materials, this technique could be a tool for biophysics studies where the potential of large effective Soret coefficient could be useful.

---

[a)] Author to whom correspondence should be addressed. Electronic mail: slan@scnu.edu.cn.

Thermophoresis, the movement of particles in a thermal gradient, or its equivalent Ludwig-Soret effect in liquids is important in many different fields like in atmospheric sciences,[1] in biology,[2] in soft-condensed matter[3] among other fields. Thus controlling and enhancing the Soret coefficient that governs the thermophoretic motion is of great interest. Almost all the studies so far utilize the nonequilibrium force generated by heating the liquid locally in the colloid[4,5] to segregate particles. However, fabrication of periodic structures is very important for photonic, plasmonic and other applications. While top-down approaches are expensive and time consuming, bottom-up approaches like formation of colloidal crystals in diluted solution[6,7] are of interest. These have been used to realize molecular sieves,[8] photonic crystals (PCs),[9] and chemical sensing.[10,11] Typically, monodispersed microspheres were self-assembled into colloidal PCs by methods that rely on different mechanisms.[12-17] Nevertheless, it is still a challenge to fabricate cost- and time-effective colloidal PC with few defects or dislocations. For example, Electrophoresis based monolayer assembly that has been reported recently,[18] depends on the charge of the particles. Similarly, an external magnetic field can also be used with suitable magnetic fluids. However, in these techniques, which depend on the external field, the orientation of the field with respect to the particles is important, leading to either clustering or linear chains. So, a suitable mechanism for fabrication of three-dimensional (3D) PCs is still challenging.

It has been shown that polysterene (PS) beads in a colloidal suspension can be assembled into two-dimensional (2D) structures by achieving large temperature gradient ($\nabla T$).[19] For this, the $\nabla T$ has to be larger than the ratio $D/aD_T$ where $a$ is the particle size and $D$ and $D_T$ are the normal diffusion coefficient and thermal diffusion coefficient of PS spheres, respectively.[20] If we introduce characteristic length which is defined as $l_T = (S_T \nabla T)^{-1}$ to describe the thermophoresis of PS spheres, then the above criterion for $\nabla T$ can be converted to a criterion for $l_T$ as follows[5,20]

$$l_T = \frac{1}{S_T \nabla T} < a. \qquad (1)$$

Here, $S_T = D_T/D$ is the Soret coefficient of PS spheres that characterizes the response of PS spheres to a thermal gradient. Physically, $l_T$ represents a length scale over which the thermal drift becomes dominant with respect to Brownian diffusion. For a given normal diffusion coefficient, a small $l_T$ implies a large thermal drift velocity $v_T$ for particles that is responsible for the crystallization of PS spheres.[19,20]

Apparently, the criterion described in equation (1) can be fulfilled by particles with a large $S_T$ in the case when a large $\nabla T$ is not easily obtained. Unfortunately, the value of $S_T$ is generally not large enough to meet the condition given in equation (1) even though it scales linearly with the size of microspheres. Recently, it has experimentally shown that the effective Soret coefficient of PS beads in a colloidal suspension can be tuned, by changing the polymer concentration in the solution, to negative values.[21] In this work we propose and demonstrate a method to achieve large Soret coefficient in colloids by suitably mixing different particles. We show that the thermophoretic motion of $Fe_3O_4$ (here after referred to as magnetic nanoparticles due to their wider application in magnetic fluids) out of the heating region results in a large effective Soret coefficient for silica beads. Consequently, silica beads are driven quickly to the heating region, forming a 3D crystal with few defects and dislocations. The binding of silica beads is so tight that a colloidal PC can be achieved after the complete evaporation of solvent, water. Thus, for fabrication of defect free 3D PCs, periodic structures for molecular sieves, among others, the proposed technique could be a low cost alternative. In addition as we use biocompatible materials, this technique could be a tool for biophysics studies where the potential of large effective Soret coefficient could be useful.

Based on the study of the thermophoresis of absorbing particles such as magnetic nanoparticles, it was found that a static temperature gradient and particle distribution can be established provided that the absorbing particles possess a positive $S_T$.[22] In this case, the thermophoretic motion of absorbing particles to cooler region provides a negative feedback to the thermal diffusion.[22-24] When we mix silica beads having negative Soret coeffient ($S_T^b$) with magnetic nanoparticles possessing positive Soret coefficient ($S_T^m$), the effective Soret coefficient for silica beads can be written as[21]

$$S_T^* = S_T^b - 2\pi(S_T^m - \frac{1}{T})a\lambda^2 c_m. \tag{2}$$

Here, $a$ is the radius of silica beads, $\lambda$ is the interaction distance between the two types of particles, and $c_m$ is the concentration (or the number density) of magnetic nanoparticles at the center of the focus in the steady state.

For silica beads used in our study, the value of $S_T^b$ is estimated to be -0.023 K$^{-1}$.[25] However, if we calculate the effective $S_T^*$ for silica beads by using $S_T^m$ = 0.15 K$^{-1}$,[26] $T$ = 300 K, $a$ = 0.8 μm, $\lambda$ = 6 nm, and $c_m$ = 1.75 x 10$^{16}$ cm$^{-3}$, it is found that a large $S_T^*$ of about 0.48 K$^{-1}$ can be achieved. This value is more than 20 times larger than that in the absence of magnetic nanoparticles. It will lead to a large $v_T$ which can be employed to assemble optical matters and PCs.

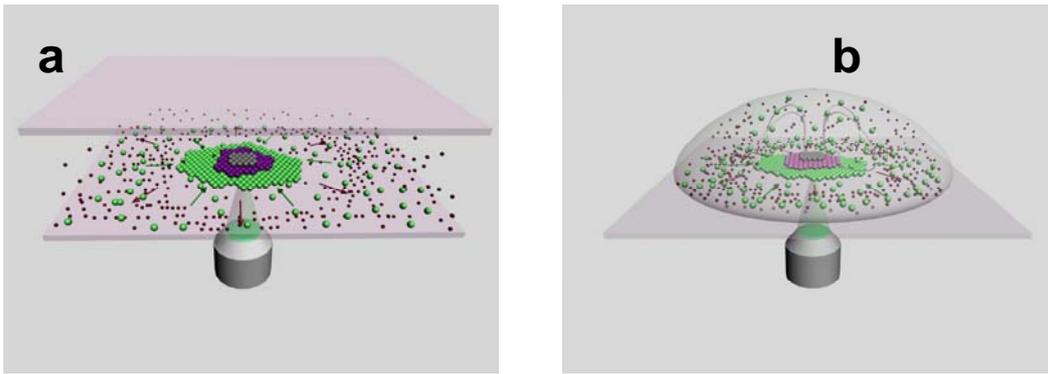

Fig. 1  Schematic showing the assembling of (a) 3D optical matter and (b) colloidal PC. The direction of movement of magnetic nanoparticles (Violet) and Silica beads (Green) are shown by arrows.

We propose and demonstrate a colloidal system on which the assembling of 3D optical matters and even colloidal PCs can be realized by using large effective Soret coefficient. A schematic of the assembling process is shown in Fig. 1. The colloidal system is composed of silica beads and magnetic nanoparticles that are uniformly distributed in water. Unlike earlier works (where the temperature gradient was produced by heating water with a focused laser beam in near infrared region), the temperature gradient that drives the thermophoresis of magnetic nanoparticles and silica beads is established by heating magnetic nanoparticles with a focused laser beam in visible region.

**Result and Discussion**

Initially, the heavier silica beads were observed to sediment on the bottom wall of the sample cell. Once the laser beam was focused into the sample cell, a fast movement of silica beads towards the focus was clearly observed. Due to the small size of magnetic nanoparticles, their thermophoretic motion could not be identified in the microscope. However, it was confirmed separately by using a pure magnetic fluid that magnetic particles possess a positive Soret coefficient in water. In this experiment, a circular depletion region surrounded by a dark ring was clearly identified when a laser beam was focused into the magnetic fluid. As predicted in Ref. 22, a steady distribution of magnetic nanoparticles was eventually achieved, resulting in a stable temperature gradient.

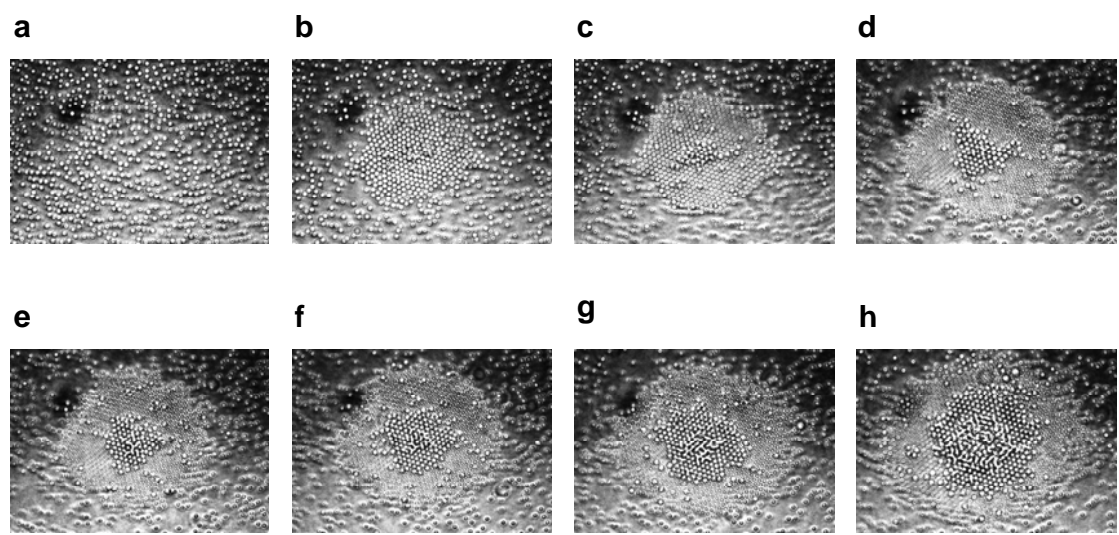

Fig. 2   CCD images showing the formation of 3D optical matter due to large nonequilibrium depletion force.

The silica beads driven by nonequilibrium depletion force, caused by the thermal diffusion of magnetic nanoparticles, moved with a large velocity to the central region to form a 2D ordered structure with a hexagonal close packed lattice. The 2D crystal grew up very quickly and more than 40 silica beads were found along the diameter of the circular-shaped crystal after about 60 seconds. Unlike the previous reports,[19,20] silica beads in the 2D crystal were regularly and closely packed into a hexagonal lattice with negligible defects. Also, it should be emphasized that a very low laser power of only about 15 mW was needed to trigger the assembly of silica beads. As mentioned above, the large drift velocity resulting from the large effective Soret coefficient is believed to be responsible for these unique features. Optical trapping effect plays a less important role in the observed assembling process as the focus of the 63x objective lens, which is similar to the diameter of silica beads, is too small to induce the assembly of a large number of silica beads. Also, the assembly did not occur when a 800-nm light is used due to weak absorption of 800 nm light in magnetic particles.

During the assembling process, a closer look at the 2D crystal revealed that some silica beads accumulated at the edge of the 2D crystal climbed up onto the crystal and moved towards the central region with a slower speed. This resulted in forming a second layer of closely packed silica beads at the central region on top of the first one. Although the area of the second layer of silica beads was much smaller than the first one, this layer-by-layer crystallization process continued. In the optimum case, a 3D crystal with five layers of regularly packed silica beads was obtained. The CCD images illustrating the assembling process of silica beads through a layer-by-layer fashion into a 3D crystal are presented in Fig. 2. The video for the entire assembling process is provided as supplement. Apart from the observation through the microscope with CCD, the laser light also acted as a probe for the ordering of the formed crystal. The evolution of the diffraction pattern of the light during the assembling process shown in Fig. 3 clearly indicates the formation of an ordered structure. The transition of the diffraction pattern from a uniform Gaussian distribution to a Debye-Scherrer ring and eventually to Bragg diffraction spots can be clearly identified.

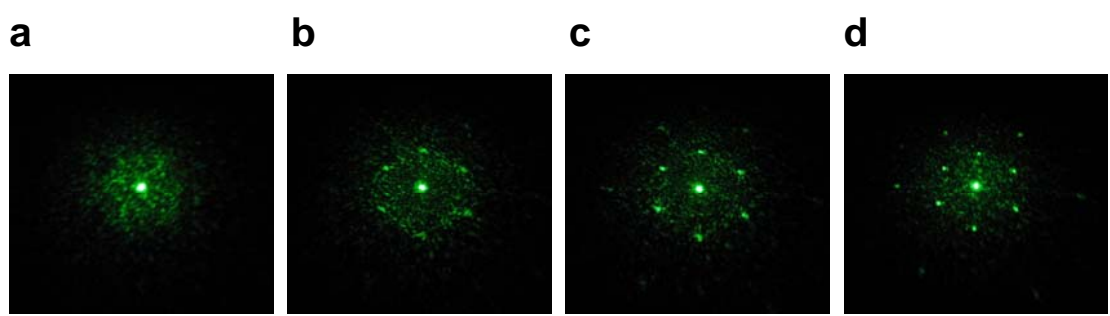

Fig. 3    Evolution of the diffraction pattern during the assembling process of the crystal.

By using colloids with different number densities of magnetic nanoparticles, it was found that the largest and best 3D crystals were obtained when the number density of magnetic particles was chosen be ~1.75 x $10^{16}$ cm$^{-3}$. In this case, the colloid was obtained by mixing the aqueous solution of silica beads with the magnetic fluid at a weight ratio of 3 : 1. This phenomenon can be interpreted by the effective Soret coefficient of silica beads that is dependent on the concentration of magnetic nanoparticles. From Eq. (2), it can be seen that the effective Soret coefficient of silica beads increases linearly with the equilibrium concentration of magnetic nanoparticles at the beam center $c_m$ that increases with the background concentration of magnetic nanoparticles. Accordingly, it was found that the assembling speed became faster with increasing concentration of magnetic nanoparticles, leading to a larger size and better quality of the formed crystal. Once the concentration of magnetic nanoparticles exceeds a certain value (i.e., $c_m$ ~ 1.75 x $10^{16}$ cm$^{-3}$ in our case), however, we observed a slow down of the assembly and a reduction in the crystal size. It implies a decrease of the effective Soret coefficient of silica beads. In fact, the effective Soret coefficient generally depends on the concentration of the particles (both magnetic and Silica beads).[27] Though the Soret coefficient can be considered as a constant when the particle concentration is low, in the case of high concentration, the interaction between particles cannot be neglected and the Soret coefficient decreases rapidly with increasing concentration.[27] Therefore, the effective Soret coefficient of silica beads begins to decrease when the concentration of magnetic nanoparticles exceeds a certain value. The 3D crystals obtained by using colloids with different number densities of magnetic particles are compared in Fig. 4.

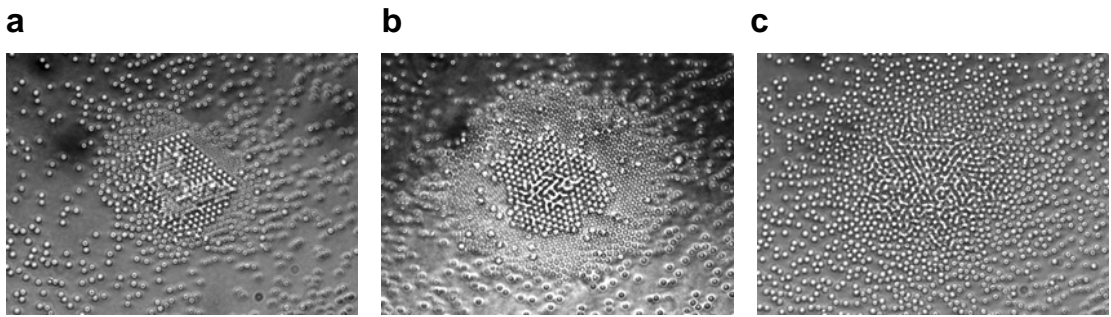

Fig. 4    3D optical matter obtained by using colloids with same number density of silica beads ($c_b$) and different number densities of magnetic nanoparticles ($c_m$). (a) $c_b$ = 3.14 x $10^9$ cm$^{-3}$, $c_m$ = 2.38 x $10^{16}$ cm$^{-3}$; (b) $c_b$ = 3.40 x $10^9$ cm$^{-3}$, $c_m$ = 1.75 x $10^{16}$ cm$^{-3}$ and (c) $c_b$ = 3.23 x $10^9$ cm$^{-3}$, $c_m$ = 4.20 x $10^{16}$ cm$^{-3}$.

Apart from the concentration of magnetic nanopartices, it is apparent from Eq. (2) that another quantity that significantly affects the effective Soret coefficient of silica beads is their own Soret coefficient ($S_T^b$). It has been known that the Soret coefficient of particles scales linearly with the size of particles. We have performed assembling experiments by using silica beads with a smaller diameter of 0.70 μm. In this case a smaller effective Soret coefficient as well as a stronger Brownian motion is expected and accordingly, it was found that only a 2D crystal with a smaller size could be obtained. Also, the assembling of PS beads with a similar diameterr (1.9 μm) but a positive Soret coefficient[21] has been carried out. Similarly, only a 2D crystal with a smaller size was achieved. In this case, the 2D crystal was formed on the top wall of the sample cell because of their lighter weight. In addition, the laser power necessary to realize the assembly was increased dramatically to about 200 mW due to their positive Soret coefficient. The 2D crystals formed by using smaller silica beads and PS beads are shown in Figs. 5(a) and 5(b), respectively.

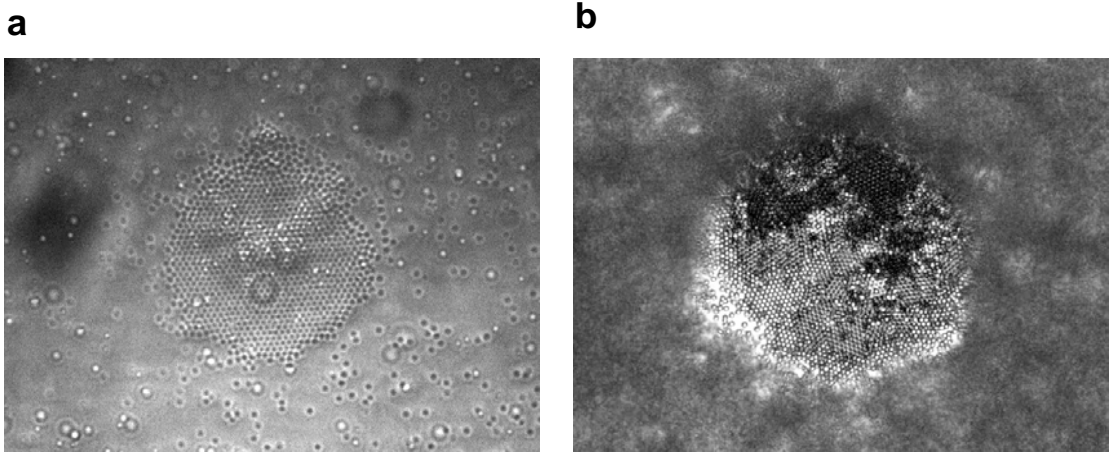

Fig. 5  2D optical mater formed by using (a) 0.7-μm silica beads and (b) 1.9-μm PS beads.

The assembling experiments described above were carried out in sample cells where the light scattering from suspended particles and walls of the cell also contribute to the formation of colloidal crystal like in optical matter. Once the laser beam is turned off, the optical matter will eventually be destroyed by the Brownian motion of particles after the disappearance of the thermal gradient. Since the characteristic length estimated from the effective Soret coefficient is quite short, it is interesting to see whether the technique can be employed to fabricate a colloidal PC or molecular sieves.

To test this idea, a droplet of the colloid containing both silica beads and magnetic nanoparticles was placed on the surface of a thin glass slide and the laser beam was focused on the surface of the glass slide. Since the height of the colloidal droplet is significantly larger than the thickness of the sample cell, the Rayleigh number which is proportional to the cubic power of the height will exceed the critical value over which Rayleigh-Bernard convection occurs.[28,29] In experiments, we did observe such a convection that drives silica beads. On the other hand, the increase of effective Soret coefficient also leads to a decrease of the threshold Rayleigh number and the appearance of Rayleigh-Bernard convection. Fortunately, the appearance of Rayleigh-Bernard convection does not affect the crystallization of silica beads. Instead, it facilitates the accumulation of silica beads around the edge of the crystal. Thus, a 3D crystal can also be created in the colloidal droplet, similar to that in sample cells.

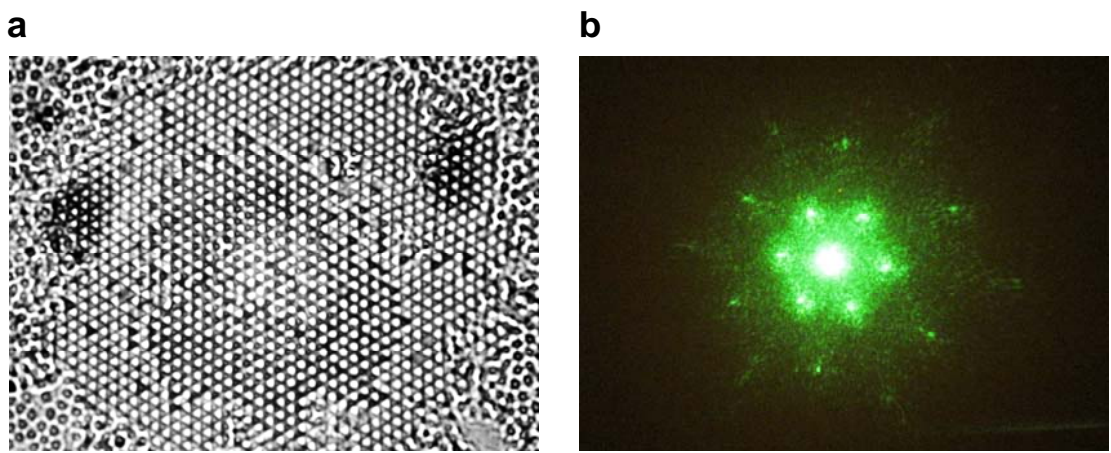

Fig. 6    CCD image (a) and the diffraction pattern (b) of the formed 3D colloidal PC.

With the formation of the 3D crystal, the evaporation of water in the colloidal droplet was accelerated because of the heating of magnetic nanoparticles. As a result, the volume of the colloidal droplet became smaller and smaller. Just before water in the droplet was completely evaporated, the silica beads in the crystal suffer from strong force that tends to break the crystal. However, the tight binding between them seems to prevent the crystal from being destroyed, leaving a perfect colloidal PC on the surface of the glass slide. The CCD image and the diffraction pattern of the formed colloidal PC are shown in Fig. 6. The colloidal PC was covered by magnetic nanoparticles after water was completely evaporated. By baking the sample at 100°C for several minutes, the magnetic-nanoparticle layer on top of the colloidal PC can be partially removed as the organic surfactant on the surface of magnetic nanoparticle cracks. The SEM images of the colloidal PC after removing the magnetic-nanoparticle layer are shown in Fig. 7. It can be seen that silica beads in the crystal are regularly arranged in a hexagonal lattice. Three layers of silica beads can be identified in the colloidal PC. Therefore, this thermophoresis-based technique offers us an easy, fast, and effective way of fabricating colloidal PCs, especially for those composed of both micro- and nanoparticles.

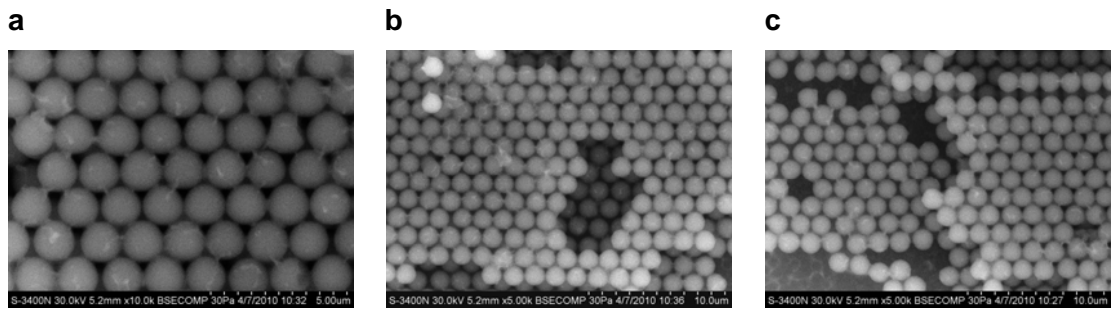

Fig. 7   SEM images of the 3D colloidal PC after removing the magnetic-nanoparticle layer.

In summary, we demonstrate a large effective Soret coefficient (more than 20 times larger) in binary colloids by suitably mixing particles having negative Soret coefficient (magnetic nanoparticles) with particles having positive Soret coefficient (Silica beads). This large effective Soret coefficient has been utilized to assemble 3D periodic structures. The large drift velocity and short characteristic length resulting from the large effective Soret coefficient lead to the formation of 3D structures with few defects and dislocations. Furthermore, the tight binding of silica beads makes it possible to obtain 3D colloidal PCs after the complete evaporation of water. This phenomenon can be employed as a low cost and faster fabrication technique for high-quality 3D optical matter and colloidal PCs. In addition, the control and enhancement of effective Soret coefficient shown in this work could be useful for various applications in widely varying fields of science.

**Methods**

The aqueous solution of silica beads used in our experiments was purchased from Duke Scientific Corporation. The diameter and weight fraction of silica beads are 1.6 μm and 2%, respectively. The water-based magnetic fluid we used was prepared by Central Iron and Steel Research Institute, China. The average diameter of magnetic particles and the weight density of the magnetic fluid were measured to be ~12 nm and 1.328 g/cm$^3$, respectively. In order to avoid aggregation, the surface of magnetic particles is capped by an organic layer. In our experiments, new colloids with uniformly distributed silica beads and magnetic particles were obtained by mixing the aqueous solution of silica beads with the magnetic fluid (or the magnetic fluid after dilution) at different weight ratios. While the number densities of silica beads ($c_b$) in these new colloids were kept to be similar, the number densities of magnetic nanoparticles ($c_m$) were deliberately made to be different in order to see their influence on the assembling process. Then, the new colloids were sonicated for half an hour and sealed into 50-μm-thick sample cells formed by two glass cover slides. The 532-nm light from a solid-state laser was employed to heat magnetic nanoparticles and to trigger the self-assembly of silica beads. The assembling process was monitored by using an inverted microscope (Zeiss Axio Observer A1) in combination with a charge-coupled device (CCD). Objective lenses with different magnifications were used and it was found that the best crystallization of silica beads was achieved by using a 63× objective lens.

# References

1. W. C. Hinds, *Aerosol Technology: Properties, Behavior, and Measurement of Airborne Particles*, Wiley Interscience **1999**.

2. L. H. Thamdrup, N. B. Larsen, A. Kristensen, *Nano Lett.* **2010**, *10*, 826.

3. R. Piazza, *Soft Matter* **2008**, *4*, 1740.

4. A. Parola, R. Piazza, *Eur. Phys. J. E* **2004**, *15*, 255.

5. R. Piazza, A. Parola, *J. Phys: Condens. Matter* **2008**, *20*, 153102.

6. N. A. Clark, A. J. Hurd, B. J. Ackerson, *Nature* **1979**, *281*, 57.

7. D. J. W. Aastuen, N. A. Clark, L. K. Cotter, *Phys. Rev. Lett.* **1986**, *57*, 1733.

8. Y. Zeng, D. J. Harrison, *Anal. Chem.* **2007**, *79*, 2289.

9. M. S. Thijssen, R. Sprik, J. E. G. J. Wijnhoven, M. Megens, T. Narayanan, A. Lagendijk, W. L. Vos, *Phys. Rev. Lett.* **1999**, *83*, 2730.

10. J. H. Holtz, S. A. Asher, *Nature* **1997**, *389*, 829.

11. R. W. J. Scott, S. M. Yang, G. chabanis, N. Coombs, D. E. Williams, G. A. Ozin, *Adv. Mater.* **2001**, *13*, 1468.

12. P. N. Pusey, W. van Megen, *Nature* **1986**, *320*, 340.

13. M. Trau, D. A. Saville, I. A. Aksay, *Science* **1996**, *272*, 706.

14. P. Sheng, W. Wen, N. Wang, H. Ma, Z. Lin, W. Y. Zhang, X. Y. Lei, Z. L. Wang, D. G. Zheng, W. Y. Tam, C. T. Chan, *Pure Appl. Chem.* **2000**, *72*, 309.

15. P. T. Korda, D. G. Grier, *J. Chem. Phys.* **2001**, *114*, 7570.

16. S. H. Park, D. Qin, Y. Xia, *Adv. Mater.* **1998**, *10*, 1028.

17. A. van Blaaderen, P. Wiltzius, *Adv. Mater.* **1997**, *9*, 833.

18. N. Aurby, P. Singh, M. Janjua, and S. Nudurupati, *PNAS* **2008**, *105*, 3711.

19. S. Duhr, D. Braun, *Appl. Phys. Lett.* **2005**, *86*, 131921.

20. F. M. Weinert, D. Braun, *Phys. Rev. Lett.* **2008**, *101*, 168301.

21. H. -R. Jiang, H. Wada, N. Yoshinaga, M. Sano, *Phys. Rev. Lett.* **2009**, *102*, 208301.

22. N. V. Tabiryan, W. Luo, *Phys. Rev. E* **1998**, *57*, 4431.

23. W. Luo, T. Du, *Phys. Rev. Lett.* **1999**, *82*, 4134.

24. M. D. Johnson, X. Duan, B. Riley, A. Bhattacharya, W. Luo, *Phy. Rev. E* **2004**, *69*, 041501.

25. N. Ghofraniha, G. Ruocco, C. Conti, *Langmuir* **2009**, *25*, 12495.

26. T. Vőlker, S. Odenbach, *Phys. Fluids* **2003**, *15*, 2198.

27. R. Piazza, A. Guarino, *Phy. Rev. Lett.* **2002**, *88*, 208302-1.

28. M. C. Cross, P. G. Daniels, P. C. Hohenberg, E. D. Siggia, *Phy. Rev. Lett.* **1980**, *45*, 898.

29. Y. Hu, R. E. Ecke, G. Ahlers, *Phy. Rev. Lett.* **1995**, *74*, 5040.